\documentclass[journal,onecolumn]{IEEEtran}

\ifCLASSINFOpdf
\else
   \usepackage[dvips]{graphicx}
\fi
\usepackage{url}

\usepackage{supertabular}
\usepackage{graphicx}
\usepackage{amsmath,empheq}
\newtheorem{theorem}{Theorem}
\newtheorem{definition}{Definition}
\newtheorem{proposition}{Proposition}

\usepackage{algorithm}
\usepackage{algorithmicx}
\usepackage{algpseudocode}
\usepackage{booktabs}
\usepackage{mathtools}
\usepackage{amssymb}
\usepackage[table]{xcolor}
\mathtoolsset{showonlyrefs} 
\definecolor{light-gray}{HTML}{FFFFFF}
\definecolor{light-cyan}{HTML}{C4C4C4}
\usepackage{float}
\newcommand{\pluseq}{\mathrel{+}=}

\usepackage{stfloats}
\usepackage{mdsymbol}
\usepackage{mathrsfs} 
\usepackage{mathtools}
\usepackage{enumitem}
\usepackage{scalerel}
\newcommand\Tau{\scalerel*{\tau}{T}}
\usepackage{verbatim}
\usepackage{longtable}

\DeclarePairedDelimiter\floor{\lfloor}{\rfloor}

\begin{document}
\title{New Classes of Binary Sequences with High Merit Factor} 
\bstctlcite{IEEEexample:BSTcontrol}

\author{Miroslav Dimitrov \IEEEmembership{Student Member, IEEE}
\thanks{This work has been partially supported by the Bulgarian National Science Fund under contract number DH 12/8, 15.12.2017.}
\thanks{M. Dimitrov  is with the Institute of Mathematics and Informatics, Bulgarian Academy of Sciences, Sofia, Bulgaria (email:mirdim@math.bas.bg).}
}%

\maketitle

\begin{abstract}
The Merit Factor (MF) measure was first introduced by Golay in 1972. Sequences possessing large values of MF are of great interest to a rich list of disciplines - from physics and chemistry to digital communications, signal processing, and cryptography.  Throughout the last half-century, manifold approaches and strategies were proposed for finding such sequences. Referenced as one of the most difficult optimization problems,  Golay wrote that it is a "challenging and charming problem". His publications on this problem spanned more than 20 years. Golay himself introduced one beneficial class of sequences, called skew-symmetric sequences, or finite binary sequences with odd lengths with alternate autocorrelation values equal to 0. Their sieving construction greatly reduces the computational efforts of finding binary sequences with odd lengths and high MF. Having this in mind, the majority of papers to be found in the literature are focused solely on this class, preferring them over binary sequences with even lengths. In this work, a new class of finite binary sequences with even lengths with alternate autocorrelation absolute values equal to 1 is presented. We show that the MF values of the new class are closely related to the MF values of adjacent classes of skew-symmetric sequences. We further introduce new sub-classes of sequences using the partition number problem and the notion of potentials, measured by helper ternary sequences. Throughout our experiments, MF records for binary sequences with many lengths less than 225, and all lengths greater than 225, are discovered. Binary sequences of all lengths, odd or even, less than $2^8$ and with MF $>8$, and all lengths, odd or even, less than $2^9$ and with MF $>7$, are now revealed.
\end{abstract}


\IEEEpeerreviewmaketitle

\section{Introduction}

Finding binary sequences whose aperiodic autocorrelation characteristics are collectively small according to some pre-defined criteria is a famous and well-studied problem. Two such measures are the Peak Sidelobe Level (PSL) and the Merit Factor (MF) value, which was first introduced by Golay in 1972 \cite{golay1972class}. Additionally, prior to Golay's definition, Littlewood  \cite{littlewood1968some} also studied the norms of polynomials with $\pm$1 coefficients on the unit circle of the complex plane, which appears to be the same problem.

 Golay's publications reveal a dedication to the merit factor problem for nearly twenty years (surveyed in \cite{jedwab2004survey}). Since then, a significant number of possible constructions of binary sequences with high merit factors were published. Near-optimal and optimal candidates are found by using heuristic search methods for longer lengths or a more direct approach, as the exhaustive search method, for smaller problem spaces.

The optimal merit factors for all binary sequences with lengths $n \leq 60$ are presented in \cite{mertens1996exhaustive}. Twenty years later, the list of optimal merit factors was extended to $n \leq 66$ \cite{packebusch2016low}. The two largest known merit factor values are 14.1 and 12.1 for lengths respectively 13 and 11. Those binary sequences are comprised of the Barker sequences \cite{barker1953group}. The merit factor problem was also attacked by various algorithms, such as the branch and bound algorithm \cite{packebusch2016low}, stochastic search algorithms like tabu search \cite{halim2008engineering}, memetic algorithm combined with tabu search \cite{gallardo2009finding}, as well as evolutionary and genetic algorithms \cite{de1992low}\cite{militzer1998evolutionary}. However, since the search space grows like $2^n$, the difficulty of finding long binary sequences with near-optimal MF significantly increases. 

A reasonable strategy for finding longer binary sequences with near-optimal merit factor is to introduce some restriction on the sequences' structure, which could decrease the size the of the search space. A well-studied restriction on the structure of the sequence has been defined by the skew-symmetric binary sequences, which were introduced by Golay \cite{golay1972class}. Having a binary sequence $(b_0,b_1, \cdots, b_{2l})$ of odd length $n = 2l+1$, the restriction is defined by $b_{l+i} = (-1)^ib_{l-i} \text{ for } i=1,2,\cdots,l.$ Golay observed that odd length Barker  \cite{barker1953group} sequences are skew-symmetric and the idea of binary sequences' sieving was proposed. This proposition was accompanied by the observation that the alternate aperiodic autocorrelation values of a skew-symmetric sequence are equal to 0. The optimal merit factors for all skew-symmetric sequences of length $n \leq 59$ were given by Golay himself \cite{golay1977sieves}. Later, the optimal merit factors for skew-symmetric sequences with lengths $n \leq 69$ \cite{golay1990new} and $n \leq 71$ \cite{de1992low} were also revealed, while the optimal skew-symmetric solutions for $n \leq 89$ and $n \leq 119$ were discovered in respectively \cite{prestwich2013improved} and \cite{packebusch2016low}.

In the case of skew-symmetric binary sequences, Golay suggested \cite{golay1975hybrid} that only one or two elements should be changed at a given optimization step. In case a new candidate has a better merit factor, the search method accepts it as a new current state and continues the optimization process. Having this in mind, a strategy of how to choose a new sequence when no acceptable neighbor sequence exists is another nuance of the search routine. 

Besides the exhaustive search approach, near-optimal results regarding skew-symmetric binary sequences' high merit factor problem are achieved by  \cite{gallardo2009finding}\cite{bovskovic2017low}\cite{brest2018heuristic}\cite{brest2020searching}. In \cite{gallardo2009finding}, the authors introduced a memetic algorithm with an efficient method to recompute the characteristics of a given binary sequence $L'$, such that $L'$ is one flip away from $L$, and assuming that some products of elements from $L$ have been already stored in memory - the tau table $\Tau(S)$ was introduced. Later, in \cite{bovskovic2017low} the principle of self-avoiding walk \cite{madras2013self} was considered. Then, in \cite{brest2018heuristic} an algorithm called xLastovka was presented and the concept of a priority queue was introduced. However, despite the rich results regarding the skew-symmetric binary sequences, the search for binary sequences with even lengths and high MF was scarcely researched. This is not surprising, since the sieving proposed by Golay is applicable to odd-length sequences only.

More details regarding the history of the problem, as well as the current achievements in the theoretical aspect of it, could be found in surveys \cite{jensen1987binary}, \cite{hoholdt1999merit} and \cite{jedwab2004survey}. Considering more theoretical results, in \cite{borwein2004binary} the authors construct an infinite family of binary sequences whose asymptotic merit factor is conjectured to be greater than 6.34. Nevertheless, as the authors stated in the abstract of their work, ``... the numerical experimentation that led to this construction is a significant part of the story.``

In this work, motivated by the absence of computationally efficient sieving for binary sequences with even length and high merit factor values, several new classes of binary sequences are proposed. In Chapter \ref{sec:problemRevisited}, we start with the definition of a class of finite binary sequences, called pseudo-skew-symmetric, with alternate auto-correlation absolute values equal to one. The class is defined by using sieving suitable for even-length binary sequences. Then, by using some mathematical observations, we show how state-of-the-art algorithms for searching skew-symmetric binary sequences with high merit factor and length $2n+1$ could be easily converted to algorithms searching pseudo-skew-symmetric binary sequences with high merit factor and lengths $2n$ or $2n+2$. More importantly, this conversion does not degrade the performance of the modified algorithm. 

In Chapter \ref{sec:partitions}, by using number partitions \cite{andrews1998theory}, additional sieving strategy for both skew-symmetric and pseudo-skew-symmetric sequences is proposed. A method of finding sub-classes of binary sequences with high MF is further discussed. The final algorithm is presented in Chapter \ref{sec:algorithm}. The experiments in Chapter \ref{sec:results} revealed that the classes defined in this work are highly promising. By using a single mid-range computer, we were able to improve all records for skew-symmetric binary sequences with lengths above 225, which were recently reached by various algorithms and a supercomputer grid. We further revealed that binary sequences with even or odd length $n$, for $n \leq 2^8$, and with merit factor strictly greater than 8, do exist, while binary sequences with even or odd length $n$, for $n \leq 2^9$ and with a merit factor strictly greater than 7 are also revealed. 

Finally, we demonstrate the efficiency of the proposed algorithm by launching it on two extremely hard search spaces of binary sequences of lengths 573 and 1009. The choice of those two specific lengths is motivated by the approximation numbers given in \cite{bovskovic2017low}, Figure 7, presented during a discussion of how much time the state-of-the-art stochastic solver lssOrel\_8 will need to reach binary sequences with the aforementioned lengths and merit factors close to  $6.34$. It was estimated that finding solutions with merit factor of 6.34 for a binary sequence with length 573 requires around 32 years, while for binary sequences with length 1009, the average runtime prediction to reach the merit factor of 6.34 is $46774481153$ years. By using the proposed in this work algorithm, we were able to reach such candidates within several hours.

\section{Preliminaries}
\label{sec:prelims}

We denote as $B=(b_0,b_1,\cdots ,b_{n-1})$ the binary sequence with length $n>1$, such that $b_i\in \{-1,1\}, 0\leq i\leq n-1$. Throughout this paper, the binary sequence could be represented with the concatenation symbol as well, i.e. $B=b_0||b_1||\cdots ||b_{n-1}$. 

The aperiodic autocorrelation function of $B$ is given by $$C_u(B)=\sum_{j=0}^{n-u-1} b_jb_{j+u}, \ \ for \ u\in \{0,1,\cdots, n-1\}.$$  

We define $C_u(B)$ for $u\in \{1, \cdots ,n-1\}$ as a sidelobe level. $C_0(B)$ is called the mainlobe. 

The enery $\mathbb{E}(B)$ of $B$ is defined as $$\mathbb{E}(B) = \sum_{u=1}^{n-1}C_u(B)^2,$$ while the merit factor, or \textbf{MF}, $\mathbb{MF}(B)$ of $B$ is defined as $$\mathbb{MF}(B) = \frac{n^2}{2\mathbb{E}(B)}.$$ 

For convenience, we will denote $C_{n-i-1}(B)$ by $\hat{C}_i(B)$. 

\section{Pseudo-Skew-Symmetric Sequences and Their Energy}
\label{sec:problemRevisited}

We consider a skew-symmetric binary sequence defined by a binary sequence $B = \left(b_0, b_1, \cdots, b_{n-1}\right)$ with an odd length $n =2l+1$.  

In this paper, for convenience, we will denote with $S_B$ the array of the sidelobes of $B$:

$$S_B= \left[\hat{C}_{0}(B), \hat{C}_{1}(B), \cdots, \hat{C}_{n-2}(B), \hat{C}_{n-1}(B)\right],$$

where $\hat{C}_{n-i-1}(B) = C_i(B)$, for $i \in \lbrace 0,1,\cdots,n-1\rbrace$. Thus,

$$\hat{C}_i(B) = C_{n-i-1}(B) = \sum_{j=0}^{n-(n-i-1)-1} b_jb_{j+(n-i-1)} =  \sum_{j=0}^{i} b_jb_{j+n-i-1}, \ \ for \ i\in \{0,1,\cdots, n-1\} .$$

The $i$-th element of a given array $S$ is denoted as $S[i]$. For practical reasons, we will define the first index of an array as 0, not 1. For example, $$S_B[0] = \hat{C}_0(B) = C_{n-1}(B).$$

Since $B$ is a skew-symmetric binary sequence, the following properties hold:

\begin{itemize}
\item{$S_B[i]=0$, for odd values of $i$.}
\item{$B[l-i] = {(-1)}^iB[l+i]$}
\end{itemize}

Having this in mind, the array of sidelobes $S_{B}$ could be represented as follows: $$S_B= \left[\hat{C}_{0}(B), 0, \hat{C}_{2}(B),0, \cdots, 0, \hat{C}_{n-3}(B), 0, \hat{C}_{n-1}(B)\right].$$

This property is beneficial for minimizing the energy $E(B)$ of the sequence. Indeed, 

$$\mathbb{E}(B) = \sum_{u=0}^{n-2}\hat{C}_u(B)^2 = \sum_{u=0,u_{even}}^{n-2}\hat{C}_u(B)^2 + \sum_{u=0,u_{odd}}^{n-2}\hat{C}_u(B)^2 = \sum_{u=0,u_{even}}^{n-2}\hat{C}_u(B)^2.$$

Unfortunately, such an approach is not suitable for binary sequences with even lengths. However, it is possible to construct an even-length binary sequence with alternating absolute values equal to 1. 

\begin{definition}[Pseudo-Skew-Symmetric Binary Sequence]\label{def:PSS}

We call a given sequence $P = a||X = Y||b$ a pseudo-skew-symmetric binary sequence, if either $X$ or $Y$ are skew-symmetric binary sequences, for some $a \in \{ -1,1 \}$ or $b \in \{ -1,1 \}$.
\end{definition}

\begin{proposition}
The sidelobes array of pseudo-skew-symmetric binary sequences consists of alternating $\pm$ ones.
\end{proposition}

\begin{IEEEproof}
Let us denote the pseudo-skew-symmetric binary sequence as $P$. By definition, $P$ could be represented as $a||A$ or $B||b$, for some skew-symmetric binary sequences $A$ or $B$.

If $P=B||b$, for some skew-symmetric binary sequence $B = \left(b_0, b_1, \cdots, b_{n-1}\right)$, then $P = \left(b_0, b_1, \cdots, b_{n-1}, b_n\right)$, where $b_n = b$. 

Thus, $$\hat{C}_i(P) = \sum_{j=0}^{i} b_jb_{j+n-i}, \ \ for \ i\in \{0,1,\cdots, n\}.$$

Therefore, the sidelobes of $P$ could be further simplified:

$$\hat{C}_i(P) = \sum_{j=0}^{i} b_jb_{j+n-i} = \sum_{j=0}^{i-1} b_jb_{j+n-i} + b_{i}b_{i+n-i} = \sum_{j=0}^{i-1} b_jb_{j+n-i} + b_{i}b_{n} = \hat{C}_{i-1}(B) + b_{i}b_{n}.$$

The last substitution arises from the definition of the sidelobe array:

$$\hat{C}_{i-1}(B) = \sum_{j=0}^{i-1} b_jb_{j+n-(i-1)-1} = \sum_{j=0}^{i-1} b_jb_{j+n-i}$$

The sidelobe array of $P$, denoted as $S_P$, could be then simplified:

\begin{equation} 
\begin{split}
S_P  & = \left[\hat{C}_{0}(P), \hat{C}_{1}(P), \cdots, \hat{C}_{n-1}(P), \hat{C}_{n}(P)\right] = \\
& =  \left[\hat{C}_{0}(P), \hat{C}_{0}(B) + b_1b_n, \hat{C}_{1}(B) + b_2b_n, \cdots, \hat{C}_{n-2}(B) + b_{n-1}b_n, \hat{C}_{n-1}(B) +b_nb_n \right] = \\
& = \left[b_0b_n, \hat{C}_{0}(B) + b_1b_n, b_2b_n, \cdots, b_{n-1}b_n, \hat{C}_{n-1}(B) +b_nb_n \right] \\
\end{split}
\end{equation}

Note that $\hat{C}_{x}(B) = 0$, for odd values of $x$. Since $b_i \in \{-1,1\}$, $\hat{C}_{x}(P) = b_xb_n = \pm 1$, for even values of $x$, which completes the proof of the first case, or more formally:

$$S_P = \left[\pm 1, \hat{C}_{0}(B) + b_1b_n, \pm 1, \cdots, \pm 1, \hat{C}_{n-1}(B) +b_nb_n \right].$$

Let us consider the second case. If $P=a||A$, then $P^{rev}=A^{rev}||a$ will posses the same sidelobes array as $P$, where $A^{rev}$ denotes the reversed version of a given binary sequence $A$. Since $A$ is skew-symmetric sequence, $A^{rev}$ is skew-symmetric as well. Thus, by applying the first case, we have:

$$S_P \equal S_{P^{rev}} =  \left[\pm 1, \hat{C}_{0}(A^{rev}) + a_1a_n, \pm 1, \cdots, \pm 1, \hat{C}_{n-1}(A^{rev}) +a_na_n \right],$$ 

where $A^{rev} = \left( a_0, a_1, \cdots, a_{n-1}, a_n \right)$ and $a_n = a$, which completes the proof.

\end{IEEEproof}

This property is beneficial for the energy $E(P)$ of the pseudo-skew-symmetric binary sequence $P$. Indeed, 

$$\mathbb{E}(P) = \sum_{u=0}^{n-1}\hat{C}_u(P)^2 = \sum_{u=0,u_{even}}^{n-1}\hat{C}_u(P)^2 + \sum_{u=0,u_{odd}}^{n-1}\hat{C}_u(P)^2 = \sum_{u=0,u_{even}}^{n-1}{\pm1}^2 + \sum_{u=0,u_{odd}}^{n-1}\hat{C}_u(P)^2 = \floor{\frac{n}{2}} + \sum_{u=1,u_{odd}}^{n}\hat{C}_u(P)^2.$$

The following property allows us to convert an existing algorithm for searching skew-symmetric binary sequences with high merit factor to an algorithm searching pseudo-skew-symmetric binary sequences and high merit factor. 

\begin{proposition}
Given a skew-symmetric binary sequence $B = \left(b_0, b_1, \cdots, b_{n-1}\right)$ with sidelobes array $$S_B= \left[\hat{C}_{0}(B), \hat{C}_{1}(B), \cdots, \hat{C}_{n-2}(B), \hat{C}_{n-1}(B)\right],$$ the following property holds: 
$$\mathbb{E}(P) = \mathbb{E}(B)+n+2b_n\delta,$$

where $P$ is the pseudo-skew-symmetric sequence $B||b_n$ and $\delta = \sum_{u=0,u_{even}}^{n-2}{\hat{C}_{u}(B)b_{u+1}}$.
\end{proposition}

\begin{IEEEproof}
Using the result from the previous proposition proof we have:
$$S_P = \left[\pm 1, \hat{C}_{0}(B) + b_1b_n, \pm 1, \cdots, \pm 1, \hat{C}_{n-1}(B) +b_nb_n \right]$$

By using the definition of energy of a binary sequence we have:
\begin{equation}
\begin{split}	
\mathbb{E}(P)-\mathbb{E}(B) = & \sum_{u=0}^{n-1}\hat{C}_u(P)^2 - \sum_{u=0}^{n-2}\hat{C}_u(B)^2 = 1 + \sum_{u=1}^{n-1}\hat{C}_u(P)^2 - \sum_{u=0}^{n-2}\hat{C}_u(B)^2 = 1 +  \sum_{u=0}^{n-2}{\hat{C}_{u+1}(P)^2 - \hat{C}_u(B)^2} = \\
& = 1 +  \sum_{u=0,u_{even}}^{n-2}{\hat{C}_{u+1}(P)^2 - \hat{C}_u(B)^2} + \sum_{u=0,u_{odd}}^{n-2}{\hat{C}_{u+1}(P)^2 - \hat{C}_u(B)^2} = \\
& = 1 + \sum_{u=0,u_{even}}^{n-2}{\left(\hat{C}_{u}(B) + b_{u+1}b_n\right)^2 - \hat{C}_u(B)^2} +  \sum_{u=0,u_{odd}}^{n-2}{\pm1^2 - 0^2}  = \\
& = 1 + \sum_{u=0,u_{even}}^{n-2}{\left(\hat{C}_{u}(B) + b_{u+1}b_n\right)^2 - \hat{C}_u(B)^2} + \floor{\frac{n-1}{2}} = \\
& = 1 + \sum_{u=0,u_{even}}^{n-2}{2\hat{C}_{u}(B)b_{u+1}b_n + \left(b_{u+1}b_n\right)^2} + \floor{\frac{n-1}{2}} = \\
& = 1 + \sum_{u=0,u_{even}}^{n-2}{2\hat{C}_{u}(B)b_{u+1}b_n} + \sum_{u=0,u_{even}}^{n-2}{\left(b_{u+1}b_n\right)^2} + \floor{\frac{n-1}{2}} = \\
& = 1 + \sum_{u=0,u_{even}}^{n-2}{2\hat{C}_{u}(B)b_{u+1}b_n} + \floor{\frac{n-1}{2}} + \floor{\frac{n-1}{2}} = n + 2b_n\sum_{u=0,u_{even}}^{n-2}{\hat{C}_{u}(B)b_{u+1}}.
\end{split}
\end{equation}

\end{IEEEproof}

The last property is of significant importance when converting an algorithm searching for skew-symmetric binary sequences, denoted as $\mathscr{A}$, to an algorithm searching for pseudo-skew-symmetric binary sequences $\mathscr{B}$ and a high merit factor. Indeed, despite the complexity of algorithm $\mathscr{A}$ we can decompose it to a tape $\cdots||\mathbb{L}_1||\cdots||\mathbb{L}_2||\cdots||\mathbb{L}_n||\cdots$, where $\mathbb{L}_i$ are stages of $\mathscr{A}$, where better candidates could be announced. They are known as local optimums in heuristic search literature. We could easily replace $\mathbb{L}_i$ with $\mathbb{L}_i||\mathbb{T}_i$, where $\mathbb{T}_i$ is a simple routine with memory and time complexity of $O(n)$, which calculates the pseudo-skew-symmetric sequences $L_i||1$ and $L_i||-1$ merit factors, where $L_i$ is the current best candidate. It should be noted that $\mathscr{B} = \cdots||\mathbb{L}_1||\mathbb{T}_1||\cdots||\mathbb{L}_2||\mathbb{T}_2||\cdots||\mathbb{L}_n||\mathbb{T}_n||\cdots$ does not interfere with the normal work of $\mathscr{A}$ by design. Furthermore, since those linear time complexity checkups are initiated on local optimums only, the delay of $\mathscr{B}$ compared to $\mathscr{A}$ caused by the additional instructions $\mathbb{T}_i$ is negligible.

We could further extend the search of highly-competitive pseudo-skew-symmetric sequences by the following observation:

\begin{proposition}
Given a skew-symmetric binary sequence $B = b_0||B'||b_{n-1}$ both binary sequences $b_0||B'$ and $B'||b_{n-1}$ are pseudo-skew-symmetric.
\end{proposition}

\begin{IEEEproof}
From the main property of the skew-symmetric sequences follows that $B'$ is skew-symmetric as well. Thus, the pseudo-skew-symmetry of $b_0||B'$ and $B'||b_{n-1}$ follows directly from definition \ref{def:PSS}. 
\end{IEEEproof}

\begin{proposition}
Given a skew-symmetric binary sequence $B = \left(b_0, b_1, \cdots, b_{n-1}\right) = b_0||B'||b_{n-1}$ with sidelobes array $$S_B= \left[\hat{C}_{0}(B), \hat{C}_{1}(B), \cdots, \hat{C}_{n-2}(B), \hat{C}_{n-1}(B)\right],$$ the following property holds: 
$$\mathbb{E}(P) = \mathbb{E}(B)+n-3+2b_{n-1}\delta,$$

where $P$ is the pseudo-skew-symmetric sequence $b_0||B'$ and $\delta = \sum_{u=1,u_{even}}^{n-2}-\hat{C}_u(B)b_u$.
\end{proposition}

\begin{IEEEproof}
We have $B = \left(b_0, b_1, \cdots, b_{n-1}\right)$ and $P = \left(b_0, b_1, \cdots, b_{n-2}\right)$. Furthermore, $$\hat{C}_i(P) = \sum_{j=0}^{i} b_jb_{j+n-2-i}, \ \ for \ i\in \{0,1,\cdots, n-2\}.$$

Decomposing $\hat{C}_i(B)$ reveals the following:

$$\hat{C}_i(B) = \sum_{j=0}^{i} b_jb_{j+n-1-i} = \sum_{j=0}^{i-1} b_jb_{j+n-1-i} + b_ib_{i+n-1-i} = b_ib_{n-1} + \hat{C}_{i-1}(P)$$ 

In other words, $\hat{C}_{i-1}(P) = \hat{C}_i(B) - b_ib_{n-1}$. Thus, by using the sidelobes array of $B$, $$S_B= \left[\hat{C}_{0}(B), 0, \hat{C}_{2}(B),0, \cdots, 0, \hat{C}_{n-3}(B), 0, \hat{C}_{n-1}(B)\right],$$ 

we could represent the sidelobes array of $P$:

\begin{equation}
\begin{split}
S_P & = \left[\hat{C}_{0}(P), \hat{C}_{1}(P), \hat{C}_{2}(P), \cdots, \hat{C}_{n-2}(P)\right] = \\
& = \left[\hat{C}_{1}(B) - b_1b_{n-1}, \hat{C}_{2}(B) - b_2b_{n-1}, \hat{C}_{3}(B) - b_3b_{n-1}, \cdots, \hat{C}_{n-1}(B) - b_{n-1}b_{n-1}\right] = \\
\end{split}
\end{equation}

By using the definition of energy of a binary sequence we have:
\begin{equation}
\begin{split}	
\mathbb{E}(P)-\mathbb{E}(B) & =  \sum_{u=0}^{n-3}\hat{C}_u(P)^2 - \sum_{u=0}^{n-2}\hat{C}_u(B)^2 = \sum_{u=0}^{n-3}\hat{C}_u(P)^2 - \left(1 + \sum_{u=1}^{n-2}\hat{C}_u(B)^2 \right)= -1 + \sum_{u=1}^{n-2}\hat{C}_{u-1}(P)^2 - \hat{C}_u(B)^2 = \\
& = -1 + \sum_{u=1,u_{even}}^{n-2}\hat{C}_{u-1}(P)^2 - \hat{C}_u(B)^2  + \sum_{u=1,u_{odd}}^{n-2}\hat{C}_{u-1}(P)^2 - \hat{C}_u(B)^2 = \\
& = -1 + \sum_{u=1,u_{even}}^{n-2}\left(\hat{C}_u(B) - b_ub_{n-1}\right)^2 - \hat{C}_u(B)^2 + \sum_{u=1,u_{odd}}^{n-2}{\pm1^2 - 0^2}= \\
& = -1 + \sum_{u=1,u_{even}}^{n-2}\left(-2\hat{C}_u(B)b_ub_{n-1} + \left(b_ub_{n-1}\right)\right)^2 + \floor{\frac{n-2}{2}} = \\
& = -1 + \sum_{u=1,u_{even}}^{n-2}-2\hat{C}_u(B)b_ub_{n-1} + \sum_{u=1,u_{even}}^{n-2}\left(b_ub_{n-1}\right)^2 + \floor{\frac{n-2}{2}} = \\
& = -1 + \sum_{u=1,u_{even}}^{n-2}-2\hat{C}_u(B)b_ub_{n-1} + \floor{\frac{n-2}{2}} + \floor{\frac{n-2}{2}} = n-3 + 2b_{n-1}\sum_{u=1,u_{even}}^{n-2}-\hat{C}_u(B)b_u
\end{split}
\end{equation}
\end{IEEEproof}

The last property further enhances the power of the algorithm. Thus now we can modify each algorithm $\mathscr{A}$, searching for skew-symmetric binary sequences with odd length $n$ and high merit factor, to an algorithm $\mathscr{B}$, searching simultaneously skew-symmetric binary sequences with odd length $n$ and pseudo-skew-symmetric binary sequences with even lengths $n-1$ and $n+1$.

\section{Restricted Skew-Symmetric Sequences Classes}
\label{sec:partitions}

Throughout the rest of this paper, we use the following definition of restriction.

\begin{definition}[Restriction Class of Binary Sequence]
We will call the class of binary sequences of length $n$, with the first $k$ elements fixed, a restriction class of order $k$ on binary sequences with length $n$. We will denote this set as $R_n^k$. If the binary sequence is skew-symmetric we will use the notation $\mathscr{R}_n^k$.
\end{definition}

It should be noted that $\mathscr{R}_n^k \subset R_n^k$. More precisely, the magnitude of $R_n^k$ is $2^{n-k}$, while the magnitude of $\mathscr{R}_n^k$ is $2^{l-k+1}$, where $n=2l+1$, since $\mathscr{R}_n^k$ is defined over the skew-symmetric binary sequences only.

A well-studied area in number theory and combinatorics is the number partition problem - distinct ways of writing a given integer number $n$ as a sum of positive integers. We define the number of possible partitions of a non-negative integer $n$ as the partition function $p(n)$. No closed-form expression for $p(n)$ is known. However, the partition functions for some different values of $n$ could be found in the online encyclopedia of integer numbers (OEIS), sequence A000041 \cite{A000041}. 

Theoretically, searching for skew-symmetric binary sequences of length $n$ with high merit factors could be parallelized to $\mid\mathscr{R}_n^k\mid$ instances. To minimize the total number of instances needed, we should consider several actions to a given skew-symmetric binary sequence $B=(b_0,b_1,\cdots ,b_{n-1})$:

\begin{itemize}
\item{Reversing $B$ defined as operator $\delta_1$: $\delta_1(B) = (b_{n-1},\cdots,b_1 ,b_0)$}
\item{Complementing $B$ defined as operator $\delta_2$: $\delta_2(B) = (\overline{b_0},\overline{b_1}, \cdots, \overline{b_{n-1}})$, where $\overline{b_i} = -b_i$}
\item{Alternating complementing of $B$ defined as operator $\delta_3$ : $\delta_3(B) = (\cdots, \overline{b_{i-2}}, b_{i-1}, \overline{b_i}, b_{i+1},  \overline{b_{i+2}}, \cdots)$}
\end{itemize}

All three operators leaves the energy of $B$ intact. If we further add the identity operator $\delta_0$ we construct a group $G$ of order 8. By using some group theory \cite{packebusch2016low}, we could derive a closed formula of the exact number of symmetry classes with length $k$: $2^{k-3} + 2^{\floor{\frac{k}{2}} - 2 + \left( k \text{ mod } 2 \right)}$. The same formula arises from the row sums of the Losanitsch's triangle (OEIS, sequence A005418 \cite{A005418}) - named after the S. Lozanić, in his work related to the symmetries exhibited by rows of paraffins \cite{losanitsch1897isomerie}. This fact could be used to partition the search space from $p(k)$ covering subsets to $2^{k-3} + 2^{\floor{\frac{k}{2}} - 2 + \left( k \text{ mod } 2 \right)}$ non-covering subsets. A similar partitioning was used in \cite{packebusch2016low} to efficiently parallelize a branch and bound algorithm for exhaustively search binary sequences with optimal merit factors. Since exhaustive search is inapplicable for large values of $n$, the following characteristic is proposed:

\begin{definition}[Potential of a Restriction Subclass]
For a skew-symemtric binary sequence $B = (b_0,b_1,\cdots ,b_{n-1})$, we fix a partitioning with length $k$: $t_0, t_1, \cdots, t_g$, s.t. $\sum_{i=0}^g{t_i}=k$. The partitioning could be projected to a skew-symmetric binary sequence with the following procedure:

$$R = \underbrace{a \cdots a}_{t_0} \underbrace{\overline{a} \cdots \overline{a}}_{t_1} \underbrace{a \cdots a}_{t_2} \underbrace{\overline{a} \cdots \overline{a}}_{t_3} \cdots \underbrace{(-1)^{g}a \cdots (-1)^{g}a}_{t_g}\underbrace{u_1u_2u_3 \cdots u_{n-2k-2}u_{n-2k-1}u_{n-2k}}_{\text{ non-fixed (free) elements}} \underbrace{f_1f_2f_3 \cdots f_{k-2}f_{k-1}f_k}_{\text{ last elements are fixed}}$$

Last $k$ elements $f_i$ are fixed due to the first $k$ elements of the sequence and its skew-symmetric property. Please note that all elements $a, \overline{a}, (-1)^ga, u_i, f_i \in \left\lbrace -1,1 \right\rbrace$. We define the potential of the binary skew-symmetric sequence $R$ as the energy of the ternary sequence $R^z$, where: 

$$R^z = \underbrace{a \cdots a}_{t_0} \underbrace{\overline{a} \cdots \overline{a}}_{t_1} \underbrace{a \cdots a}_{t_2} \underbrace{\overline{a} \cdots \overline{a}}_{t_3} \cdots \underbrace{(-1)^{g}a \cdots (-1)^{g}a}_{t_g}\underbrace{000 \cdots 000}_{n-2k \text{ zeroed elements}} \underbrace{f_1f_2f_3 \cdots f_{k-2}f_{k-1}f_k}_{\text{ last elements are fixed}}$$
\end{definition}

$R^z$ is ternary since we have introduced a new element $0$. This way we could not only focus on the complete sidelobes of $R$ but take under consideration the non-complete fragments of sidelobes of $R$, where the fixed elements of the sequence play a role. For example, let us consider a skew-symmetric binary sequence $Q$ with length $n=21$, a restriction $k=6$ and a partition $1,1,2,2$:

$$Q = \underbrace{a}_{1} \underbrace{\overline{a}}_{1} \underbrace{aa}_{2} \underbrace{\overline{a}\overline{a}}_{2} \underbrace{u_1u_2u_3 \cdots u_{9}}_{\text{ non-fixed (free) elements}} \underbrace{f_1f_2f_3f_4f_5f_6}_{\text{elements are fixed}}$$

Since $Q$ is skew-symmetric we know that $Q[l-i]=(-1)^i Q[l+i]$, for $n=2l+1$. If we take $i=l$ we have $Q[0]=(-1)^lQ[n-1]$. In the current example, $n=21$ and $l=10$. Therefore $f_6 = Q[20] = Q[0](-1)^l= Q[0]$. By following the same routine we could reveal all values of $f_i$:

$$Q = \underbrace{a}_{1} \underbrace{\overline{a}}_{1} \underbrace{aa}_{2} \underbrace{\overline{a}\overline{a}}_{2} \underbrace{u_1u_2u_3 \cdots u_{9}}_{\text{ non-fixed (free) elements}} \underbrace{a}_{1} \underbrace{\overline{a}\overline{a}}_{2} \underbrace{aaa}_{3}$$

We could easily derive $Q^z$:

$$Q^z = \underbrace{a}_{1} \underbrace{\overline{a}}_{1} \underbrace{aa}_{2} \underbrace{\overline{a}\overline{a}}_{2} \underbrace{000000000}_{9} \underbrace{a}_{1} \underbrace{\overline{a}\overline{a}}_{2} \underbrace{aaa}_{3}$$

Without loss of generality, let us fix $a=1$. Then, we have $\overline{a} = -1$, and $$Q^z = (1,-1,1,1,-1,-1,0,0,0,0,0,0,0,0,0,1,-1,-1,1,1,1)$$

Thus, the potential of the partition $1,1,2,2$ is equal to $\mathbb{E}(Q^z)$. The sidelobes' array of $Q^z$ is $$S_{Q^z} = \left[ 1,  0,  1,  0,  1,  0, -5,  0,  3,  0, -1,  0,  0,  0,  0,  0,  0, 0, -4,  0\right],$$

therefore $\mathbb{E}(Q^z) = \sum_u{S_{Q^z}}_u^2 = 54.$ The cardinality of the set $\mathscr{R}_{21}^6$ is $\mid \mathscr{R}_{21}^6 \mid = 2^{6-3} + 2^{\floor{\frac{6}{2}} - 2 + \left( 6 \text{ mod } 2 \right)} = 2^3 + 2^1 = 10$. A list of unique partitions in $\mathscr{R}_{21}^6$ could be find in Table \ref{tab:partitions621}. For simplicity, we denote as $\mathscr{R}_{n}^{k \mid m}$ those partitions of size $k$ over $n$, which posses exactly $m$ elements. For example, referring Table \ref{tab:partitions621}, the partition $(6)$ is in $\mathscr{R}_{21}^{6 \mid 1}$, partitions $(5,1)$, $(4,2)$ and $(3,3)$ are in $\mathscr{R}_{21}^{6 \mid 2}$, partitions $(4,1,1)$, $(3,1,2)$, $(3,2,1)$ and $(2,2,2)$ are in $\mathscr{R}_{21}^{6 \mid 3}$, while partitions $(2,1,2,1)$ and $(2,1,1,2)$ are in $\mathscr{R}_{21}^{6 \mid 4}$. 

\begin{table}
\begin{center}
\caption{A list of unique partitions in $\mathbb{R}_{21}^6$}
\label{tab:partitions621}
\ttfamily
\rowcolors{2}{light-gray}{light-cyan}
\begin{tabular}{lc}
{Partition} & {$\pm$ Notation}   \\
\toprule
$6$ & ['+', '+', '+', '+', '+', '+']\\
$5,1$ & ['+', '+', '+', '+', '+', '-']\\
$4,1,1$ & ['+', '+', '+', '+', '-', '+']\\
$4,2$ & ['+', '+', '+', '+', '-', '-']\\
$3,1,2$ & ['+', '+', '+', '-', '+', '+']\\
$3,2,1$ & ['+', '+', '+', '-', '-', '+']\\
$3,3$ & ['+', '+', '+', '-', '-', '-']\\
$2,1,2,1$ & ['+', '+', '-', '+', '+', '-']\\
$2,1,1,2$ & ['+', '+', '-', '+', '-', '-']\\
$2,2,2$ & ['+', '+', '-', '+', '-', '-']\\
\showrowcolors
\bottomrule
\end{tabular}
\end{center}
\end{table}

Given a partition $t_0, t_1, \cdots, t_{g}$ of size $k$, we will denote the set of skew-symmetric binary sequences defined by the partition as $\mathbb{B}_{n}^{t_0, t_1, \cdots, t_{g}}$. Please note that $\mathbb{B}_{n}^{t_0, t_1, \cdots, t_{g}} \subset \mathscr{R}_{n}^{k \mid g+1} \subset \mathscr{R}_{n}^{k}$. Finally, the potential of a given partition set $S$ is denoted as $\mathscr{U}(S)$. 

Few remarks regarding the sidelobes of a given potential ternary sequence should be made. In case we are interested in the potential of $\mathscr{R}_{n}^{k}$, the sidelobes of the ternary sequence could be divided into three distinct sections:

\begin{itemize}
\item{\textbf{Head}: first $2k$ sidelobes. The first $k$ sidelobes are shared among all sequences in this class, i.e. they are immutable. The next $k$ sidelobes are just partially revealed.}
\item{\textbf{Body}: the mid $n-3k$ sidelobes. They are all equal to zero.}
\item{\textbf{Tail}: the last $k$ sidelobes. Ignoring the sidelobes equal to zero, the remaining sidelobes are even numbers greater than 0. They are not shared among the sequences in this class, i.e. they are mutable. However, partial information about their final value is gathered.}
\end{itemize}

The actual calculation of the potential $\mathscr{U}(\mathscr{R}_{n}^{k})$ gives an equal priority to the value of the elements in the head and the tail. However, we could tweak the actual energy calculation while minimizing the energy of the elements to prefer minimizing the elements in the head more, than minimizing the elements in the tail. All elements inside the tail are even numbers.  If we prefer to minimize the energy of the summands, rather than minimizing their overall sums, we could normalize the tail by dividing its sidelobes values by 2. We will define this value as a normalized potential, denoted as $\mathscr{U}^{\star}(\mathscr{R}_{n}^{k})$.

As a final remark, please note that despite $\mathscr{R}_{n}^{k}, \mathscr{R}_{n+1}^{k}, \mathscr{R}_{n+2}^{k}, \cdots$ is an infinite sequence of non-intersecting finite sets, their potentials and normalized potentials are equal. More formally, $\forall i \geq n \forall j \geq i: \mathscr{U}(\mathscr{R}_{i}^{k})=\mathscr{U}(\mathscr{R}_{j}^{k}) \& \mathscr{U}^{\star}(\mathscr{R}_{i}^{k})=\mathscr{U}^{\star}(\mathscr{R}_{j}^{k})$.

During our research, by using an exhaustive search, we have calculated all the potentials, as well as normalized potentials, of set partitions of the form $\mathscr{R}_{n}^{k \mid g}$, for $38 < k < 115$ and some values of $g \in \left[ 4, 12\right]$. For speeding up the exhaustive routine, the following restriction of the partitions were further applied: $\forall i: t_i \geq t_{i+1}$. As an illustration, various partitions having an optimal potential and normalized potential are given in Table \ref{tab:potentials}. 

\begin{table}
\begin{center}
\caption{Some partitions with optimal and normalized potentials}
\label{tab:potentials}
\ttfamily
\rowcolors{2}{light-gray}{light-cyan}
\begin{tabular}{l|c|c|c|c}
{Class} & {$\mathscr{U}$ optimal} & {$\mathscr{U}$} & {$\mathscr{U}^{\star}$ optimal} & {$\mathscr{U}^{\star}$} \\
\toprule
$\mathscr{R}_{n}^{39 \mid 4}$ & 18,11,6,4 & 3731 & 18,11,6,4 & 1082\\
$\mathscr{R}_{n}^{41 \mid 6}$ & 17,9,6,4,3,2 & 2217 & 17,9,6,4,3,2 & 813\\
$\mathscr{R}_{n}^{47 \mid 9}$ & 18,8,5,4,3,3,2,2,2 & 1859 & 11,9,5,5,5,3,3,3,3 & 830\\
$\mathscr{R}_{n}^{56 \mid 4}$ & 27,14,9,6 & 12856 & 27,14,9,6 & 3472\\
$\mathscr{R}_{n}^{68 \mid 7}$ & 25,11,10,7,5,5,5 & 9596 & 25,12,9,8,6,4,4 & 3040\\
$\mathscr{R}_{n}^{79 \mid 9}$ & 26,12,10,7,6,6,6,4,2 & 11667 & 28,14,10,7,6,6,4,2,2 & 3702\\
\showrowcolors
\bottomrule
\end{tabular}
\end{center}
\end{table}

\section{Algorithm for Finding Binary Sequences with Arbitrary Length and High Merit Factor}
\label{sec:algorithm}

In our previous work, we were able to reduce the memory complexity of state-of-the-art algorithms from $O\left(n^2\right)$ to $O\left(n\right)$, by keeping the time complexity linear as well. This was achieved by the usage of the following theorem:

\begin{theorem}
\label{theorem:meritFactor}
Given two skew-symmetric sequences $B$ and $B^q$ with length $n=2l+1$, where $B^q$ corresponds to $B$ with $q$-th and $n-q-1$-th bit flipped for some fixed $q<l$, and with sidelobe array of $B$ denoted as $S$, the following property holds:
\begin{equation} 
\label{theorem1:property3}
\begin{split}
& \mathbb{E}(B^q) = \mathbb{E}(B) + \sum_{r=q+1, r \neq 2q+1}^{n-q-1}(16+\sigma\kappa\epsilon_1) +  \sum_{r=n-q, r \neq 2q+1}^{n-1} (\kappa(\epsilon_2+\sigma\epsilon_1) + 32 + 32\sigma\epsilon_1\epsilon_2) + \sum_{r \geq n-q, r \leq n-1, r = 2q+1}(16+\kappa\epsilon_2),
\end{split}
\end{equation}

where $\sigma=(-1)^{l-q}$, $\kappa = -8S_rL[q]$, $\epsilon_1(r) = B[r-q-1]$, $\epsilon_2(r) = B[q+r-n]$.
\end{theorem}

The proof of Theorem \ref{theorem:meritFactor} could be found in \cite{dimitrov2021skew}.

By achieving both linear time and memory complexities, we can utilize all the threads of a given central processing unit. Furthermore, the memory requirements of a given algorithm are significantly reduced.

\algrenewcommand\algorithmicindent{0.5em}%
\begin{algorithm}[]
\caption{Algorithm for searching skew-symmetric and pseudo-skew-symmetric binary sequences with arbitrary lengths and high merit factors.  }
\label{algor:AlgoMF}
\begin{algorithmic}[1]
\Procedure{MF}{$n, t_0, t_1, \cdots, t_g, \mathbb{T}_i, \mathbb{T}_o, \mathbb{T}_a$}

\State $\bot_{n-1}$, $\bot_{n}$, $\bot_{n+1}$, $w_o \gets 0, 0, 0, 0$ 
\While {True}	 
	\State $\mathbb{H}, w_i, \gets \left\lbrace \emptyset \right\rbrace, 0$
	\State $B \gets$ random , s.t. $B \in \mathbb{B}_n^{t_0,t_1, \cdots, t_g} \subset \mathscr{R}_{n}^{k},$ for $k = \sum_{i=0}^g{t_i}$.
	\State $\mathbb{H}$.add(hash$(B)$)
	\State $V \gets \mathbb{E}(B)$   
	\While {True}	
		\State bestN $\gets$ pickBetterNeighborIndex$(B)$
		\If{bestN $== -1$}
			\State break
		\EndIf
		\State Flip(bestN, $B$)
		\State $V \gets \mathbb{E}(B)$ 
		\State $w_i \pluseq 1$
		\State $\mathbb{H}$.add(hash($B$))
		\If{$\frac{n^2}{2V} > \bot_n$}
			\State $\bot_n \gets \frac{n^2}{2V}$		
		\EndIf
		\If{$\frac{n^2}{2V} \geq \mathbb{T}_a$}
		 \If{$\frac{{(n+1)}^2}{2(V+n+2b_n\delta)} > \bot_{n+1}$  }
		 	\State $\bot_{n+1} \gets \frac{{(n+1)}^2}{2(V+n+2b_n\delta)}$		
		 \EndIf
		 \If{$\frac{{(n-1)}^2}{2(V+n-3+2b_{n-1}\delta)} > \bot_{n-1} $}
		 	\State $\bot_{n-1} \gets \frac{{(n-1)}^2}{2(V+n-3+2b_{n-1}\delta)}$		
		 \EndIf
		\EndIf
		\If{$w_i > \mathbb{T}_i$}
			\State $w_o \pluseq 1$
			\State break
		\EndIf
	\EndWhile
\If{$w_o > \mathbb{T}_o$}
	\State break
\EndIf
\EndWhile
\EndProcedure
\end{algorithmic}
\end{algorithm}

In Algorithm \ref{algor:AlgoMF} a pseudo-code of the proposed routine is presented. The following additional notations and remarks should be considered:

\begin{itemize}
\item{$n$ - an odd integer number}
\item{$t_0, t_1, \cdots, t_g$ - the partition search space to search through.}
\item{$\mathbb{T}_i$ - an inner threshold value. When the inner counter $w_i$ reaches $\mathbb{T}_i$, the set is flushed and the whole routine restarts. The threshold value $\mathbb{T}_i$ constrains the size of the set $\mathbb{H}$.}
\item{$\mathbb{T}_o$ - an outer threshold value. When the outer counter $w_o$ reaches $\mathbb{T}_o$, the program is terminated. }
\item{$\mathbb{T}_a$ - an activator threshold value. For example, the probability of finding pseudo-skew-symmetric sequence with length $n-1$ or $n+1$ and merit factor $X$, from a skew-symmetric sequence with length $n$ and merit factor $X-1$, is negligible for higher values of $X$. Thus, we could save time and efforts to repeatedly probe the adjacent pseudo-skew-symmetric sequences.}
\item{$\bot_{n-1}, \bot_n, \bot_{n+1}$ - the best candidates found, in terms of merit factor value, for respectively pseudo and not pseudo-skew-symmetric sequences of lengths $n-1,n$ and $n+1$.}

\item{$\mathbb{H}$ - a set of hashes of the visited candidates. We make sure to avoid already visited nodes.}

\item{$\mathbb{H}$.add(hash$(B)$) - adding the hash of the binary sequence $B$ to the set $\mathbb{H}$.}
\item{pickBetterNeighborIndex - a function, which returns the index of a better unexplored neighbor of $B$, i.e. the binary sequence with a distance of exactly 1 flip away from $B$, s.t. its hash does not belong to the set $\mathbb{H}$. An optimized derivative-based pseudo-code of this helper function is discussed in our previous work \cite{dimitrov2021skew}.}
\end{itemize}

\section{Results}
\label{sec:results}

Algorithm \ref{algor:AlgoMF} was implemented (C++) on a general-purpose computer equipped with a central processing unit with 8 cores and 16 threads. Despite using just a single low-budget personal computer, we were able to improve all the results, for all skew-symmetric lengths in the range 225-451, announced in literature and reached by using a supercomputer grid. Furthermore, by using classes of pseudo-skew-symmetric sequences, we were able to simultaneously reach binary sequences of even lengths between 225 and 512, and beyond, with merit factors greater than 7. We demonstrate the efficiency of our approach by publishing a complete list of binary sequences, for both even and odd lengths up to $2^8$, and merit factors greater than 8. The list is further accompanied by a complete list of binary sequences, for both even and odd lengths up to $2^9$, and merit factors greater than 7 (see Tables \ref{tab:records1} - \ref{tab:records11}).

We further demonstrate the power and efficiency of the proposed algorithm by launching it on binary sequences of lengths 573 and 1009. As mentioned earlier, the choice of those two specific lengths is motivated by the approximation numbers given in \cite{bovskovic2017low}, Figure 7, presented during a discussion of how much time the state-of-the-art stochastic solver lssOrel\_8 will need to reach binary sequences with the aforementioned lengths and merit factors close to  $6.34$. It was estimated that finding solutions with a merit factor of 6.34 for a binary sequence with length 573 requires around 32 years, while for binary sequences with length 1009, the average runtime prediction to reach the merit factor of 6.34 is $46774481153$ years. By using the proposed in this work algorithm, we were able to reach such candidates within several hours (see Table \ref{tab:records12}). By further applying some operators on the skew-symmetric binary sequence of length 1009 found, several sequences of lengths 1006, 1007, 1008, and 1010 with MF greater than 6.34 were also revealed. The same argument is true for the other sequence of length 573, but since the results are too many we omit the data.

For convenience, we denote the operators acting on binary sequences as shown in Table \ref{tab:operators}. Please note that operator $\eta_0$ activated on a given skew-symmetric binary sequence $a||L||b$ will yield another skew-symmetric binary sequence $L$, while all other operators activated on the same skew-symmetric binary sequence will yield a pseudo-skew-symmetric sequence. Throughout the tables with reported records, the classes denoted with $\Omega$ represent the best-known result to be found in the literature for the current length (all in Table \ref{tab:records1}, for the lengths between 172 and 226). It should be emphasized, that all records achieved by starting from a sequence of class $\Omega$, are directly calculated without the usage of any additional stochastic routine. All other records throughout the tables (classes $\mathbb{B}$) are achieved by using a heuristic search. All sequences are presented in hexadecimal format with zeroes omitted. It should be noted, that as soon as the algorithm finds a record sequence of a given length, it automatically continues to the next search space. In some cases, we required a little bit more demanding goal, i.e. MF greater than 8 (for sequences with lengths less than about 256), or MF greater than 7 (for sequences with lengths less than about 512). Some records were found for several minutes, while others required a little bit more effort of several hours.  

\begin{table}[H]
\begin{center}
\caption{A list of used operators acting on binary sequences}
\label{tab:operators}
\ttfamily
\rowcolors{2}{light-gray}{light-cyan}
\begin{tabular}{ll}
{Operator} & {Action}   \\
\toprule
$\eta_0$ & $a||L||b \circ \eta_0 = L$\\
$\eta_1$ & $L \circ \eta_1 = L||1$\\
$\eta_2$ & $L \circ \eta_2 = L||-1$\\
$\eta_3$ & $a||L \circ \eta_3 = L$\\
$\eta_4$ & $L||b \circ \eta_4 = L$\\
$\eta_5$ & $L \circ \eta_5 = 1||L$\\
$\eta_6$ & $L \circ \eta_6 = -1||L$\\

\showrowcolors
\bottomrule
\end{tabular}
\end{center}
\end{table}

\section{Conclusions}
In this work, several new sieving classes of binary sequences with high merit factors are proposed. A class of finite binary sequences of even length called pseudo-skew-symmetric with alternate absolute values equal to 1 is found. We show how algorithms searching for skew-symmetric binary sequences with high MF of odd length, could be easily transformed to search pseudo-skew-symmetric binary sequences of even length. We have further introduced the concept of potentials measured by number partitioning and helper ternary sequences. This allows us to reach binary sequences of any length with record merit factors in record times. Numerical pieces of evidence suggest that there are various non-intersecting classes of binary sequences with asymptotic values greater than 7, and even 8, which is significantly greater than the current largest known asymptotic merit factor of approximately 6.34. 





\bibliographystyle{IEEEtran}
\bibliography{refs}

\newpage
\begin{table}
\begin{center}
\caption{A list of binary sequences with record merit factor values}
\label{tab:records1}
\ttfamily
\rowcolors{2}{light-gray}{light-cyan}
\begin{tabular}{llp{9cm}cl}
{$n$} & Class & {Record sequence in HEX} & Old MF & New MF\\
\toprule
$172$ & $\Omega_{173}^{} \circ \eta_4$ &  fe03184fe780309206b6663e571d355a6ac59356a & 8.8363 & 9.052631578947368\\
$178$ & $\Omega_{179}^{} \circ \eta_4$ &  2c7bd3a7034ccbfe886e8a084688550ccf2b613d16c & 8.2125 & 9.29149560117302\\
$180$ & $\Omega_{179}^{} \circ \eta_1$ &  b1ef4e9c0d332ffa21ba28211a2154333cad84f45b0 & 8.5353 & 8.553326293558607\\
$182$ & $\Omega_{183} \circ \eta_4 $ & 3cfe712191dc7d1c57c81ec5a8d7edbd6ddb9a3b654d2 & 8.2194 & 8.928301886792452\\ 
$184$ & $\Omega_{183} \circ \eta_2 $ & f3f9c4864771f4715f207b16a35fb6f5b76e68ed9534a & 8.2980 & 8.636734693877552\\ 
$186$ & $\Omega_{185} \circ \eta_1 $ & 233da5ef19ed00149bc0c4644d4b9c1550e992e878b375 & 8.3606 & 8.627431421446383\\ 
$190$ & $\Omega_{187} \circ \eta_0 \circ \eta_0 \circ \eta_4 $ & 190c8d2692191f17dba56aff75407e11d7b5b9a3863c8cb9 & 8.5021 & 9.195109526235354\\
\bottomrule
\end{tabular}
\end{center}
\end{table}


\newpage
\begin{table}
\begin{center}
\caption{A list of binary sequences with record merit factor values}
\label{tab:records2}
\ttfamily
\rowcolors{2}{light-gray}{light-cyan}
\begin{tabular}{llp{9cm}cl}
{$n$} & Class & {Record sequence in HEX} & Old MF & New MF\\
\toprule

$244$ & $\mathbb{B}_{243}^{} \circ \eta_1$ &  fffff181ce703f081fc72c1e7498cd9cc98f65ac36d5a8b52b64d a9b55555 & 7.1730 & 8.51974813967\\
$245$ & $\mathbb{B}_{243}^{} \circ \eta_1 \circ \eta_5$ &  1fffff181ce703f081fc72c1e7498cd9cc98f65ac36d5a8b52b64 da9b55555 & 7.3237 & 8.5799028016\\
$246$ & $\mathbb{B}_{243}^{} \circ \eta_1 \circ \eta_1 \circ \eta_5$ &  3ffffe3039ce07e103f8e583ce9319b39931ecb586dab516a56c9 b536aaaab & 7.1650 & 8.33324153126\\
$247$ & $\mathbb{B}_{243}^{} \circ \eta_1 \circ \eta_1 \circ \eta_2 \circ \eta_5$ &  7ffffc60739c0fc207f1cb079d2633673263d96b0db56a2d4ad93 6a6d55556 & 7.109 & 8.22889128675\\
$248$ & $\mathbb{B}_{243}^{} \circ \eta_1 \circ \eta_1 \circ \eta_2 \circ \eta_5 \circ \eta_6$ &  7ffffc60739c0fc207f1cb079d2633673263d96b0db56a2d4ad93 6a6d55556 & - & 8.01668404588\\
$249$ & $\mathbb{B}_{249}^{16}$ &  1ffffc212fe40e94e19e33d2972665b1b1e663783d3259a4f84ae 543a2d5555 & 8.1323 & 8.20119047619\\
$250$ & $\mathbb{B}_{249}^{} \circ \eta_2$ &  3ffff8425fc81d29c33c67a52e4ccb6363ccc6f07a64b349f095c a8745aaaaa & 7.1988 & 8.19564647259\\
\bottomrule
\end{tabular}
\end{center}
\end{table}


\newpage
\begin{table}
\begin{center}
\caption{A list of binary sequences with record merit factor values}
\label{tab:records3}
\ttfamily
\rowcolors{2}{light-gray}{light-cyan}
\begin{tabular}{lp{3cm}p{9cm}cl}
{$n$} & Class & {Record sequence in HEX} & Old MF & New MF\\
\toprule
$279$ & $\mathbb{B}_{275}^{} \circ \eta_1 \circ \eta_2 \circ \eta_1 \circ \eta_5$ &  7fffff001fb1c8c7e463d23c792731e0f123b4a5b36396d23d26e 56c8db15aab555555 & - & 7.05209277043\\
$280$ & $\mathbb{B}_{281}^{} \circ \eta_4$ &  fffffc007936c187ec0dd8f03db13cda71b39278cb138b52d88d4 ea594e31a554aaaaa & - & 7.27002967359\\
$281$ & $\mathbb{B}_{281}^{22,11}$ &  1fffff800f26d830fd81bb1e07b6279b4e36724f1962716a5b11a 9d4b29c634aa955555 & 7.5058 & 7.7050156128\\
$282$ & $\mathbb{B}_{281}^{} \circ \eta_2$ &  3fffff001e4db061fb03763c0f6c4f369c6ce49e32c4e2d4b6235 3a96538c69552aaaaa & - & 7.38933283776\\
$283$ & $\mathbb{B}_{283}^{22,11}$ &  7ffffe003f9c0f3c38d867139330e53e47a17a46b06d331b12658 db4b2d49ab556aaaaa & 7.5088 & 8.17067945317\\
$284$ & $\mathbb{B}_{283}^{} \circ \eta_1$ &  fffffc007f381e7871b0ce272661ca7c8f42f48d60da663624cb1 b6965a9356aad55555 & - & 7.89815902859\\
$285$ & $\mathbb{B}_{285}^{22,11}$ &  1fffff800ff218fc31e07b06d24f26c6c6d999c6c6c634e3c6b16 a5b2d49a354aa955555 & 7.0142 & 7.46827877896\\
\bottomrule
\end{tabular}
\end{center}
\end{table}


\newpage
\begin{table}
\begin{center}
\caption{A list of binary sequences with record merit factor values}
\label{tab:records4}
\ttfamily
\rowcolors{2}{light-gray}{light-cyan}
\begin{tabular}{lp{3cm}p{9cm}cl}
{$n$} & Class & {Record sequence in HEX} & Old MF & New MF\\
\toprule

$313$ & $\mathbb{B}_{313}^{24,11,9,4}$ &  1fffffe003fe04b878f0b32666cda7c2c5a4f98f4994e1ec2d61c c666330b49690ea552aa555555 & 7.5547 & 7.62048848787\\
$314$ & $\mathbb{B}_{313}^{} \circ \eta_1$ &  3fffffc007fc0970f1e1664ccd9b4f858b49f31e9329c3d85ac39 8ccc661692d21d4aa554aaaaab & - & 7.31315828512\\
$315$ & $\mathbb{B}_{315}^{23,10,7,4}$ &  7fffff003f87c0f184fe339cf0c6179e38de666466668db69a164 d2c9b36ac592d4a5ab552aaaaa & 7.4661 & 7.5204638472\\
$316$ & $\mathbb{B}_{315}^{} \circ \eta_2$ &  fffffe007f0f81e309fc6739e18c2f3c71bcccc8cccd1b6d342c9 a59366d58b25a94b56aa555554 & - & 7.25908694388\\
$317$ & $\mathbb{B}_{317}^{}$ &  1ffffff003f920d9f36093b4ec686396db0f225e25e234b1c7926 86c4f138a7359ca3952ab555555 & - & 7.46131571132\\
$318$ & $\mathbb{B}_{317}^{} \circ \eta_2$ &  3fffffe007f241b3e6c12769d8d0c72db61e44bc4bc469638f24d 0d89e2714e6b39472a556aaaaaa & - & 7.23658222413\\
$319$ & $\mathbb{B}_{319}^{25,10,7}$ &  7fffffc00fe0674e1939ce09ccf92963cd83c378da34b58cb61f1 acc9d6c9b196c2656ad54aaaaaa & 7.4224 & 7.45720357614\\
\bottomrule
\end{tabular}
\end{center}
\end{table}


\newpage
\begin{table}
\begin{center}
\caption{A list of binary sequences with record merit factor values}
\label{tab:records5}
\ttfamily
\rowcolors{2}{light-gray}{light-cyan}
\begin{tabular}{lp{3cm}p{9cm}cl}
{$n$} & Class & {Record sequence in HEX} & Old MF & New MF\\
\toprule

$346$ & $\mathbb{B}_{339}^{} \circ \eta_1 \circ \eta_1 \circ \eta_1 \circ \eta_1 \circ \eta_2 \circ \eta_5 \circ \eta_6$ &  1ffffffc001ff8463d86f06b4ce1f0b4e9399b1b6c372c71b1993 84f0b5a4cf06b469d26e955aaad555555e & - & 7.10818192614\\
$347$ & $\mathbb{B}_{347}^{26,13,10,0}$ &  7fffffe000ffcc0f1e19c327c33966f03f21d8e17a6367a16d897 2b52e61b34a73499692d4caad556aaaaaa & 7.1698 & 7.23177177177\\
$348$ & $\mathbb{B}_{347}^{} \circ \eta_2$ &  ffffffc001ff981e3c33864f8672cde07e43b1c2f4c6cf42db12e 56a5cc36694e6932d25a9955aaad555554 & - & 7.13888233907\\
$349$ & $\mathbb{B}_{349}^{24,9,9,9}$ &  1fffffe00ff8033e76c49b0a79274c36a5ad0f0399b3931992b4b c1e072cf63960b18ec76532a954aa555555 & 7.1295 & 7.24488460623\\
$350$ & $\mathbb{B}_{349}^{} \circ \eta_2$ &  3fffffc01ff0067ced893614f24e986d4b5a1e073367263325697 83c0e59ec72c1631d8eca6552a954aaaaaa & - & 7.14619064287\\
$351$ & $\mathbb{B}_{349}^{} \circ \eta_2 \circ \eta_1$ &  7fffff803fe00cf9db126c29e49d30da96b43c0e66ce4c664ad2f 0781cb3d8e582c63b1d94caa552a9555555 & 7.0911 & 7.14043120436\\
$352$ & $\mathbb{B}_{349}^{} \circ \eta_2 \circ \eta_1 \circ \eta_5$ &  ffffff803fe00cf9db126c29e49d30da96b43c0e66ce4c664ad2f 0781cb3d8e582c63b1d94caa552a9555555 & - & 7.05603644647\\
\bottomrule
\end{tabular}
\end{center}
\end{table}


\newpage
\begin{table}
\begin{center}
\caption{A list of binary sequences with record merit factor values}
\label{tab:records6}
\ttfamily
\rowcolors{2}{light-gray}{light-cyan}
\begin{tabular}{lp{3cm}p{9cm}cl}
{$n$} & Class & {Record sequence in HEX} & Old MF & New MF\\
\toprule

$381$ & $\mathbb{B}_{381}^{27,13,10,5}$ &  1ffffffc001ff818f0787c0d9e13e0cf0e49e43cc39c667333666 d92cd2e58e4b4ca53a59cad696b49a955aaad555555 & 7.106 & 7.2018753721\\
$382$ & $\mathbb{B}_{381}^{} \circ \eta_2$ &  3ffffff8003ff031e0f0f81b3c27c19e1c93c8798738cce666ccd b259a5cb1c96994a74b395ad2d69352ab555aaaaaaa & - & 7.08437712399\\
$383$ & $\mathbb{B}_{383}^{27,13,10,5}$ &  7ffffff0007fe07319cce478e730f318f01bc1e3963c3f27272b4 b61b694b952d932d326da46cc993256aa5552aaaaaa & 7.0314 & 7.07480466866\\
$384$ & $\mathbb{B}_{383}^{} \circ \eta_1 \circ \eta_2 \circ \eta_3$ &  ffffffc001ff81cc673391e39cc3cc63c06f078e58f0fc9c9cad2 d86da52e54b64cb4c9b691b3264c95aa9554aaaaaaa & - & 7.08923076923\\
$385$ & $\mathbb{B}_{385}^{27,13,10,7}$ &  1ffffffc001ff80d8c72d86c73343b439960f721f8d2d61b331a7 c3c95a374a7992f12f336c69c36c9ca955aaad555555 & 7.0772 & 7.24887519562\\
$386$ & $\mathbb{B}_{385}^{} \circ \eta_2$ &  3ffffff8003ff01b18e5b0d8e668768732c1ee43f1a5ac366634f 8792b46e94f325e25e66d8d386d93952ab555aaaaaaa & - & 7.1296774811\\
$387$ & $\mathbb{B}_{387}^{27,13,10,5}$ &  7ffffff0007fe033c1f66139c2d327c96786f078d8dc39b33339b 48d8da52e5a61ca730f49b166294b356aa5552aaaaaa & 7.1502 & 7.19974040958\\
\bottomrule
\end{tabular}
\end{center}
\end{table}


\newpage
\begin{table}
\begin{center}
\caption{A list of binary sequences with record merit factor values}
\label{tab:records7}
\ttfamily
\rowcolors{2}{light-gray}{light-cyan}
\begin{tabular}{lp{3cm}p{9cm}cl}
{$n$} & Class & {Record sequence in HEX} & Old MF & New MF\\
\toprule

$415$ & $\mathbb{B}_{415}^{28,14,10,6}$ &  7ffffff8001ff80f0799f00f6c378793c96331a730e43694ecd33 0cec1e346d32793361cb1a5a34e2d5299a52d5aa9555aaaaaaa & - & 7.25891427126\\
$416$ & $\mathbb{B}_{415}^{} \circ \eta_1$ &  fffffff0003ff01e0f33e01ed86f0f2792c6634e61c86d29d9a66 19d83c68da64f266c39634b469c5aa5334a5ab552aab5555555 & - & 7.04855001629\\
$417$ & $\mathbb{B}_{417}^{28,13,12,8,6,6,4}$ &  1ffffffe000fff00fc0fc92787993c3e0793934f867264730c739 936cb36e636694f39396a52d39969638d4ad4ab554aaa5555555 & - & 7.22610538564\\
$418$ & $\mathbb{B}_{417}^{} \circ \eta_2$ &  3ffffffc001ffe01f81f924f0f32787c0f27269f0ce4c8e618e73 26d966dcc6cd29e7272d4a5a732d2c71a95a956aa9554aaaaaaa & - & 7.13683522588\\
$419$ & $\mathbb{B}_{417}^{} \circ \eta_2 \circ \eta_5$ &  7ffffffc001ffe01f81f924f0f32787c0f27269f0ce4c8e618e73 26d966dcc6cd29e7272d4a5a732d2c71a95a956aa9554aaaaaaa & - & 7.05120893244\\
$420$ & $\mathbb{B}_{421}^{} \circ \eta_4$ &  fffffff0007ff807e07f031b2d86c9cc670cd8786d23c9c399b0c d399b49cb70e5a58cd264c9ce58f39352a56a55aaa5552aaaaaa & - & 7.04585397028\\
$421$ & $\mathbb{B}_{421}^{28,13,12,8,6,6,4}$ &  1ffffffe000fff00fc0fe06365b0d9398ce19b0f0da4793873361 9a73369396e1cb4b19a4c9939cb1e726a54ad4ab554aaa5555555 & - & 7.25327385824\\
\bottomrule
\end{tabular}
\end{center}
\end{table}


\newpage
\begin{table}
\begin{center}
\caption{A list of binary sequences with record merit factor values}
\label{tab:records8}
\ttfamily
\rowcolors{2}{light-gray}{light-cyan}
\begin{tabular}{lp{3cm}p{9cm}cl}
{$n$} & Class & {Record sequence in HEX} & Old MF & New MF\\
\toprule

$442$ & $\mathbb{B}_{441}^{} \circ \eta_2$ &  3ffffff0007fe0479c679c39c3873b1ce523c4e1f09784e46c3f0 cdc8cd2b4e46c5a1d296c4b706c93b25b49b49a649a456aa5552a aaaaa & - & 7.07994491556\\
$443$ & $\mathbb{B}_{441}^{} \circ \eta_2 \circ \eta_1$ &  7fffffe000ffc08f38cf3873870e7639ca4789c3e12f09c8d87e1 9b919a569c8d8b43a52d896e0d92764b6936934c9348ad54aaa55 55555 & - & 7.06694274397\\
$444$ & $\mathbb{B}_{441}^{} \circ \eta_2 \circ \eta_1 \circ \eta_2$ &  ffffffc001ff811e719e70e70e1cec73948f1387c25e1391b0fc3 372334ad391b16874a5b12dc1b24ec96d26d269926915aa9554aa aaaaa & - & 7.07798362775\\
$445$ & $\mathbb{B}_{445}^{27,13,10,7,3,3}$ &  1ffffffc001ff80e1cce3c34c6783d2663e14b913b461ec631e07 98736996a5b26c5a6f13b90fa52663d2966cf2d24cda4a955aaad 555555 & - & 7.05317709075\\
$446$ & $\mathbb{B}_{445}^{} \circ \eta_2$ &  3ffffff8003ff01c399c78698cf07a4cc7c29722768c3d8c63c0f 30e6d32d4b64d8b4de27721f4a4cc7a52cd9e5a499b4952ab555a aaaaaa & - & 7.12296784359\\
$447$ & $\mathbb{B}_{445}^{} \circ \eta_2 \circ \eta_5$ &  7ffffff8003ff01c399c78698cf07a4cc7c29722768c3d8c63c0f 30e6d32d4b64d8b4de27721f4a4cc7a52cd9e5a499b4952ab555a aaaaaa & - & 7.19410239793\\
$448$ & $\mathbb{B}_{445}^{} \circ \eta_2 \circ \eta_2 \circ \eta_5$ &  fffffff0007fe0387338f0d319e0f4998f852e44ed187b18c781e 61cda65a96c9b169bc4ee43e94998f4a59b3cb4933692a556aab5 555554 & - & 7.168\\
\bottomrule
\end{tabular}
\end{center}
\end{table}


\newpage
\begin{table}
\begin{center}
\caption{A list of binary sequences with record merit factor values}
\label{tab:records9}
\ttfamily
\rowcolors{2}{light-gray}{light-cyan}
\begin{tabular}{lp{3cm}p{9cm}cl}
{$n$} & Class & {Record sequence in HEX} & Old MF & New MF\\
\toprule

$465$ & $\mathbb{B}_{461}^{} \circ \eta_1 \circ \eta_1 \circ \eta_1 \circ \eta_1$ &  1ffffffe0007ff00f0f633781fc1a59a58dacc6e70f394e6d1a6c 6c3348f32c6c61bc64f934b646cc1c9e19e1ad5a973274b4ab556 aaa5555555f & - & 7.08469855832\\
$466$ & $\mathbb{B}_{461}^{} \circ \eta_1 \circ \eta_1 \circ \eta_1 \circ \eta_1 \circ \eta_2$ &  3ffffffc000ffe01e1ec66f03f834b34b1b598dce1e729cda34d8 d86691e658d8c378c9f2696c8d98393c33c35ab52e64e96956aad 554aaaaaabe & - & 7.09057663423\\
$467$ & $\mathbb{B}_{467}^{30,15,12,10,6,6,6}$ &  7ffffffe0003ffc00fc0fc1e0c39e19ce1c725e1399933cc8db83 6691e6691e635b8dccb3199b16872496c9969b4d694ad4ad54aab 5556aaaaaaa & - & 7.0075509286\\
$468$ & $\mathbb{B}_{469}^{} \circ \eta_4$ &  fffffffe0003ffc03f03f19c781f870e6631e8c63ccc78f138f1c c9c96ce1c9cc92db12da4ccb64de93666d25a95a4992b52b54aab 5556aaaaaaa & - & 7.16795392067\\
$469$ & $\mathbb{B}_{469}^{}$ &  1fffffffc0007ff807e07e338f03f0e1ccc63d18c7998f1e271e3 99392d9c3939925b625b49996c9bd26ccda4b52b493256a56a955 6aaad5555555 & - & 7.3743127263\\
$470$ & $\mathbb{B}_{469}^{} \circ \eta_1$ &  3fffffff8000fff00fc0fc671e07e1c3998c7a318f331e3c4e3c7 32725b38727324b6c4b69332d937a4d99b496a569264ad4ad52aa d555aaaaaaab & - & 7.25165780316\\
$471$ & $\mathbb{B}_{469}^{} \circ \eta_1 \circ \eta_2$ &  7fffffff0001ffe01f81f8ce3c0fc3873318f4631e663c789c78e 64e4b670e4e6496d896d2665b26f49b33692d4ad24c95a95aa555 aaab55555556 & - & 7.12719270064\\
\bottomrule
\end{tabular}
\end{center}
\end{table}


\newpage
\begin{table}
\begin{center}
\caption{A list of binary sequences with record merit factor values}
\label{tab:records10}
\ttfamily
\rowcolors{2}{light-gray}{light-cyan}
\begin{tabular}{lp{3cm}p{9cm}cl}
{$n$} & Class & {Record sequence in HEX} & Old MF & New MF\\
\toprule

$486$ & $\mathbb{B}_{487}^{} \circ \eta_4$ &  3fffffff0007ff807e07cf038f8c79c1e60f8799b0dc9f30f3391 b0e730f48f4b364b1b9334b358dcb199694a65ad96c9492b4d6a5 6a9556aab5555555 & - & 7.14230420321\\
$487$ & $\mathbb{B}_{487}^{30,13,12,8,6,6,5}$ &  7ffffffe000fff00fc0f9e071f18f383cc1f0f3361b93e61e6723 61ce61e91e966c9637266966b1b96332d294cb5b2d9292569ad4a d52aad556aaaaaaa & - & 7.19784522003\\
$488$ & $\mathbb{B}_{487}^{} \circ \eta_1$ &  fffffffc001ffe01f81f3c0e3e31e707983e1e66c3727cc3cce46 c39cc3d23d2cd92c6e4cd2cd6372c665a52996b65b2524ad35a95 aa555aaad5555555 & - & 7.19033816425\\
$489$ & $\mathbb{B}_{489}^{30,15,12,8,6,4}$ &  1fffffff8000fff00fc2c7e07878e1ec1f8c6d927c99626679339 86d24f264e634e3c6993396662798d639c6c95ac5a49696a56c2d 4ab554aaa95555555 & - & 7.20070464948\\
$490$ & $\mathbb{B}_{489}^{} \circ \eta_2$ &  3fffffff0001ffe01f858fc0f0f1c3d83f18db24f932c4ccf2673 0da49e4c9cc69c78d32672ccc4f31ac738d92b58b492d2d4ad85a 956aa95552aaaaaaa & - & 7.09305760709\\
$491$ & $\mathbb{B}_{489}^{} \circ \eta_2 \circ \eta_1$ &  7ffffffe0003ffc03f0b1f81e1e387b07e31b649f2658999e4ce6 1b493c99398d38f1a64ce59989e6358e71b256b16925a5a95b0b5 2ad552aaa55555555 & - & 7.08519955328\\
$492$ & $\mathbb{B}_{495}^{} \circ \eta_0 \circ \eta_4$ &  fffffff0001ffc01f81f06d836cc2fc30d9c3b659b4e4ed38d8da 4e72666961e66726c78d8db0ec6c39863b498d34af4ce358e5295 a954aa95552aaaaaa & - & 7.07045215563\\
\bottomrule
\end{tabular}
\end{center}
\end{table}


\newpage
\begin{table}
\begin{center}
\caption{A list of binary sequences with record merit factor values}
\label{tab:records11}
\ttfamily
\rowcolors{2}{light-gray}{light-cyan}
\begin{tabular}{lp{3cm}p{9cm}cl}
{$n$} & Class & {Record sequence in HEX} & Old MF & New MF\\
\toprule

$506$ & $\mathbb{B}_{501}^{} \circ \eta_2 \circ \eta_2 \circ \eta_1 \circ \eta_1 \circ \eta_5$ &  3fffffff8000fff00fc1e4f30f865f207e0f34da63598c5e0d939 bc360e4d8ccc9ce4a72d1939ca5ec99f261cf34a56a35e694b34e 5ad4ab554aaa955555553 & - & 7.05994595489\\
$507$ & $\mathbb{B}_{507}^{29,15,11,9,6,6,4}$ &  7ffffffc0007ff007e07c1b30db303f0e1c9c6d99996e58f4f0e7 863948cf183592cdc1b65a6d2c2d86e19998e49c96d2b5338d339 4a56a552aa5554aaaaaaa & - & 7.23144657627\\
$508$ & $\mathbb{B}_{507}^{} \circ \eta_1$ &  fffffff8000ffe00fc0f83661b6607e1c3938db3332dcb1e9e1cf 0c72919e306b259b836cb4da585b0dc33331c9392da56a671a672 94ad4aa554aaa95555555 & - & 7.10293955741\\
$509$ & $\mathbb{B}_{509}^{26,13,12,9,7,6,6,4,2}$ &  1ffffff8003ffc01fc0fc3078f266781f81f393278cda4cb1c78d 2649c6963ccccd2786d8e63c96db0ce1cc9633935a95a96663496 b2d4ad5aad552aa9555555 & - & 7.03489193005\\
$510$ & $\mathbb{B}_{511}^{} \circ \eta_4$ &  3ffffffe0007ffc007f01fc079b43c5b0ce72761e437872d2cccb 4e4db61b18ccc9b1a71ce4f0ccc3c36972e5a76364cb1ed2f196a d5ab56aad556aaa5555555 & - & 7.00889248181\\
$511$ & $\mathbb{B}_{511}^{27,15,11,11,7,7,7,7}$ &  7ffffff0001ffc007f01fc079258d264c3613c61b1b1c8d8665a5 8c78c6c4b4b3c3c4e4da4d8786658dc9393964b1634c670d871a5 4a952a554aa95552aaaaaa & - & 7.22166602135\\
$512$ & $\mathbb{B}_{511}^{} \circ \eta_1$ &  ffffffe0003ff800fe03f80f24b1a4c986c278c3636391b0ccb4b 18f18d8969678789c9b49b0f0ccb1b927272c962c698ce1b0e34a 952a54aa9552aaa5555555 & - & 7.09417622862\\
\bottomrule
\end{tabular}
\end{center}
\end{table}


\newpage
\begin{table}
\begin{center}
\caption{A list of binary sequences with record merit factor values}
\label{tab:records12}
\ttfamily
\rowcolors{2}{light-gray}{light-cyan}
\begin{tabular}{lp{3cm}p{9cm}cl}
{$n$} & Class & {Record sequence in HEX} & Old MF & New MF\\
\toprule

$573$ & $\mathbb{B}_{573}^{27,15,11,11,7,7,7,7}$ &  1ffffffc0007ff001fc07f01e7c2787b0e1e5a7e3cf0f64bc39ce 372d9b399ac9b26e333333246318c199319c3724d92d0e74b4d25 61e5a4b16962d65ab56ad5aab556aaad55555 & - & 6.82937432399\\                 
$1006$ & $\mathbb{B}_{1009}^{} \circ \eta_0 \circ \eta_4$ &  3fffffffff000007fff0001ff801fe01e03fc0be03e9c60fb70e1 f039b0e3c762c73ca479ccc6f0d9c9327c97879938d8dcd937330 e1793867d8761e19f1b59f1b59a5a769d669397a4b33739cdc9c9 3996978d6338d9cb46ccd96e0d36c276d24b192b5a4b714a6d852 a50ad52a5aa55aa955aaab5556aaaab55555555 & - & 6.35677047348\\                  $1007$ & $\mathbb{B}_{1009}^{} \circ \eta_0$ &  7ffffffffe00000fffe0003ff003fc03c07f817c07d38c1f6e1c3 e07361c78ec58e7948f3998de1b39264f92f0f3271b1b9b26e661 c2f270cfb0ec3c33e36b3e36b34b4ed3acd272f49666e739b9392 732d2f1ac671b3968d99b2dc1a6d84eda4963256b496e294db0a5 4a15aa54b54ab552ab5556aaad55556aaaaaaaa & - & 6.41941303825\\                               $1008$ & $\mathbb{B}_{1009}^{} \circ \eta_4$ &  fffffffffe00000fffe0003ff003fc03c07f817c07d38c1f6e1c3 e07361c78ec58e7948f3998de1b39264f92f0f3271b1b9b26e661 c2f270cfb0ec3c33e36b3e36b34b4ed3acd272f49666e739b9392 732d2f1ac671b3968d99b2dc1a6d84eda4963256b496e294db0a5 4a15aa54b54ab552ab5556aaad55556aaaaaaaa & - & 6.41811107180\\                                $1009$ & $\mathbb{B}_{1009}^{39,21,15,15,10,10,8,8,4}$ &  1fffffffffc00001fffc0007fe007f80780ff02f80fa7183edc38 7c0e6c38f1d8b1cf291e7331bc36724c9f25e1e64e3637364dccc 385e4e19f61d87867c6d67c6d66969da759a4e5e92ccdce737272 4e65a5e358ce3672d1b3365b834db09db492c64ad692dc529b614 a942b54a96a956aa556aaad555aaaaad55555555 & - & 6.41690827959\\
\bottomrule
\end{tabular}
\end{center}
\end{table}

\end{document}